\pdfoutput=1
\RequirePackage{ifpdf}
\ifpdf % We are running pdfTeX in pdf mode
\documentclass[pdftex]{sigma}
\else
\documentclass{sigma}
\fi

\numberwithin{equation}{section}

\begin{document}
%\allowdisplaybreaks

\newcommand{\arXivNumber}{1904.09323}

\renewcommand{\PaperNumber}{073}

\FirstPageHeading

\ShortArticleName{A K\"ahler Compatible Moyal Deformation of the First Heavenly Equation}

\ArticleName{A K\"ahler Compatible Moyal Deformation \\ of the First Heavenly Equation}

\Author{Marco MACEDA and Daniel MART\'INEZ-CARBAJAL}
\AuthorNameForHeading{M.~Maceda and D.~Mart\'inez-Carbajal}
\Address{Departamento de F\'isica, Universidad Aut\'onoma Metropolitana, Av. San Rafael Atlixco 186,\\ C.P.~03340, Deleg. Iztapalapa, Mexico City, M\'exico}
\Email{\href{mailto:mmac@xanum.uam.mx}{mmac@xanum.uam.mx}, \href{mailto:danielmc@xanum.uam.mx}{danielmc@xanum.uam.mx}}

\ArticleDates{Received June 07, 2019, in final form September 08, 2019; Published online September 22, 2019}

\Abstract{We construct a noncommutative K\"ahler manifold based on a non-linear perturbations of Moyal integrable deformations of $D=4$ self-dual gravity. The deformed K\"ahler manifold preserves all the properties of the commutative one, and we obtain the associated noncommutative K\"ahler potential using the Moyal deformed gravity approach. We apply this construction to the Atiyah--Hitchin metric and its K\"ahler potential, which is useful in the description of interactions among magnetic monopoles at low energies.}

\Keywords{heavenly equations; Moyal deformation; Atiyah--Hitchin metric}

\Classification{37K10; 53C26; 53D55; 70H06; 83C20}

\section{Introduction}

Several applications of hyper-K\"ahler manifolds in four dimensions involve gravitational instantons, non-linear graviton theory and the heavenly equations~\cite{Eguchi1980213,gibbons1979,Ko198151,Penrose1976,Penrose:1976jq}; they have also been extensively used in supersymmetric field theories~\cite{AlvarezGaume:1981hm,hitchin1987}. Of particular interest is their appearance in topological field theories and string theory, where in some cases the moduli spaces have a hyper-K\"ahler structure. The existence of such a structure provides a more profound and alternative understanding of a physical system in general.

Even though hyper-K\"ahler manifolds have been analysed in great detail, there exist examples where the metric has proved to be difficult or impossible to calculate. Nevertheless, an algebraic description of four-dimensional non-compact hyper-K\"ahler manifolds possessing one abelian isometry was given in~\cite{Bakas:1996gf}. In that work, a self-duality condition for the Killing vector associated to the isometry plays a fundamental role in the analysis and classification manifolds; the translational or rotational character of the isometry translates into the existence of a~translational or rotational Killing vector.

Two particular examples of self-dual manifolds with rotational Killing symmetries are the Eguchi--Hanson and the Taub-NUT metrics; both metrics share ${\rm SO}(3)\times {\rm SO}(2)$ as a larger group of isometries, but these two group factors act differently on each spacetime. ${\rm SO}(3)$ is a~translation symmetry for the Eguchi--Hanson metric with ${\rm SO}(2)$ acting as a rotational one; the situation is the opposite for the Taub-NUT spacetime. On the other hand, a relevant example of a spacetime admitting only a rotational isometry is the Atiyah--Hitchin (AH) metric that arises on the moduli space $M_{2}^{0}$ of BPS ${\rm SU}(2)$ monopoles~\cite{TheGeometryandDynamicsofMagneticMonopoles,GIBBONS1986183}.

As noted in a series of works~\cite{1985AtiyahandN.J.Hitchin, TheGeometryandDynamicsofMagneticMonopoles,GIBBONS1986183}, the AH spacetime is a useful tool in the description of interactions among magnetic monopoles. More specifically, the AH metric is the metric on the moduli space of charge-two non-abelian magnetic ${\rm SU}(2)$ monopole with a fixed centre. Its geodesics describe low-energy monopoles interacting through the exchange of massless photons and scalars~\cite{TheGeometryandDynamicsofMagneticMonopoles}; at long distances, it reduces to the Taub-NUT space with negative mass parameter~\cite{Hanany:2000fw}. The structure of the AH metric is a four-dimensional hyper-K\"ahler manifold with~${\rm SO}(3)$ isometry; the ${\rm SO}(3)$ group does not rotate the three K\"ahler forms, and it is the only specific example of a four-dimensional hyper-K\"ahler space without tri-holomorphic isometries. Furthermore, the AH spacetime is a self-dual solution to Einstein field equations~\cite{Atiyah425} and all of its Killing vectors lack a self-dual covariant derivative~\cite{gibbons1988}; the isometry group ${\rm SU}(2)$ is often identified with a supersymmetry group~\cite{GIBBONS1988679}.

In general, the metric related to self-dual vacuum solutions to Einstein's field equations is a solution to the first, or the second, heavenly equations~\cite{Plebanski1975}. The original motivation for heavenly equations, heavenly metrics and heavenly spaces was the desire to obtain real solutions to Einstein's equations on real manifolds and spawned several papers on the subject and its generalisations~\cite{Alexandrov:2009vj,Boyer:1983ip, Esposito:2005zt,Finley:1976gr,Husain:1993dp}. Assuming that the anti-self-dual part of the Weyl tensor was algebraically special, the equivalence between the vacuum Einstein field equations in complex spacetime and the heavenly equation was established in~\cite{PhysRevLett.37.493}. The heavenly equations are also integrable using the twistor formalism~\cite{doi:10.1063/1.526652, Penrose1976} and several examples are known~\cite{GarciaD.1977,J.F.PlebanskiandG.F.TorresdelCastillodoi:10.1063/1.525498}; they are a constant source of research in mathematics and physics.

It is of interest to consider integrable generalisations of the heavenly equation due to its applications to physical systems; its modifications may allow a description of new phenomena or interactions not present in the standard description of a system. For instance, one of the possible generalisations that may be relevant is the one related to the Moyal $\star$-product. Within the context of particles and fields, the Moyal product provides a straightforward generalisation to noncommutative field theories and introduces, for example, new interactions in the Standard Model that may be probed in the laboratory. Several applications in quantum gravity also exist, in particular regarding the issue of the singularities and the thermodynamical properties of solutions in general relativity.

In the case of the first heavenly equation, a Moyal deformation of this equation has been done independently by Strachan~\cite{STRACHAN199263} and Takasaki~\cite{TAKASAKI1994111}. Furthermore, a suitable deformed differential calculus was introduced in~\cite{Strachan:1996gx} to deal with the deformation. When applied to integrable systems, the equations
\begin{gather*}
{\rm d}\widehat{\boldsymbol{\Omega}} = 0, \qquad
\widehat{\boldsymbol{\Omega}} \wedge \widehat{\boldsymbol{\Omega}} = 0,
\end{gather*}
for a 2-form $\widehat{\boldsymbol{\Omega}}$ provide a concise writing of the integrability conditions for the deformed system.

We want to analyse the consequences of a noncommutative structure on the moduli space of interacting magnetic monopoles; we expect that such a deformation gives rise to interactions that may be identified with some already known or produces new ones. For the classical case, the K\"ahler potential for the AH metric was obtained by Olivier~\cite{Olivier1991} following an approach based on the existence of a $\eta$-self-dual Killing vector in conjunction with previous results obtained by Boyer and Finley~\cite{C.P.BoyerandJ.D.FinleyIIIdoi:10.1063/1.525479}. In our approach, we consider a Moyal deformation of this K\"ahler potential, and we require that the deformed potential $\widehat{\Omega}$ must share the same features and properties of the classical one.

The plan of the present paper is as follows. In Section~\ref{sec:MOYAL-DEFORMATION-OFTHE FIRST HEAVENLY EQUATION}, we review the noncommutative deformation of the Monge--Amp\`ere or first heavenly equation using the Moyal $\star$-product; in this section, we write the equation satisfied by the noncommutative contributions that preserve the first heavenly equation. In Section~\ref{sec:Integrable-Systems}, we also recall the anti-self-dual vacuum Einstein equations that determine the structure of complex four-dimensional metrics of Euclidean signature. How integrability is preserved in the Moyal-deformed case for these spaces and the conditions under which we guarantee that the Moyal-deformed potential is K\"ahler are presented there.

Afterwards, in Section~\ref{sec:Noncommutative-gravity}, we use the Moyal deformed gravity approach to rewrite the deformed K\"ahler potential $\widehat{\Omega}$ in terms of local frame fields $\hat{e}_{\mu}^{a}(x, \theta)$; we also obtain general expressions for the corresponding deformed metric elements. We discuss the deformed K\"ahler potential up to first order on the noncommutative parameter $\theta$ in greater detail in Section~\ref{sec:The-Deformed-K=0000E4hler Potential}; through an integration procedure, we explicitly specify the first order modifications to the deformed metric and vierbien. The results in Sections~\ref{sec:Noncommutative-gravity} and~\ref{sec:The-Deformed-K=0000E4hler Potential} are put together in Section~\ref{sec:Deformed--self-dual Riemman 4- metric} to analyse the case of the AH metric. There, we obtain the Moyal deformed AH metric as a function of the original AH metric, the corrections to the K\"ahler manifold and the complex coordinates used in the original formulation by Olivier. We finally end with our conclusions and some remarks on future work.

\section{Moyal deformation of the first heavenly equation}\label{sec:MOYAL-DEFORMATION-OFTHE FIRST HEAVENLY EQUATION}

It is well known that the first and second heavenly equations describe the general metric of a self-dual vacuum space-time~\cite{Plebanski1975} and that these equations are integrable by twistor methods~\cite{doi:10.1063/1.526652,Penrose1976,1980grg2.conf..283P}. Pleba\'nski~\cite{Plebanski1975} showed that complex metrics with a self-dual Riemann tensor, to be referred to as self-dual metrics,
can be described in terms of one function $\Omega$, the K\"ahler potential, satisfying the first heavenly equation
\begin{gather}
\left\{ \Omega_{p}, \Omega_{q}\right\} _{\rm PB}=\Omega_{p\bar{p}}\Omega_{q\bar{q}}-\Omega_{p\bar{q}}\Omega_{q\bar{p}}=1.
\label{eq:Monge-Amper=0000E9 equation-1-1}
\end{gather}
This equation is also called Pleba\'nski's first equation; it defines a completely multidimensional integrable system with an infinite number of conservation laws, hierarchy and Lax pair formulation~\cite{Newman1978,Park7e5835d7735e46e5bb42f99940967dc6,PARKdoi:10.1142/S0217751X92000624, IABStrachan0264-9381-10-7-017,Strachan:1994ts}. The K\"ahler potential $\Omega$ is an unknown function of suitable spacetime coordinates $x^\mu := (p, q, \bar{p}, \bar{q})$; it gives a local expression of self-dual vacuum Einstein spaces.

In this section we outline the procedure in~\cite{STRACHAN199263} to obtain the deformation of the first heavenly equation using Strachan's idea. The starting point is to replace the Poisson bracket
\begin{gather*}
\{F, G\} _{\rm PB}=\frac{\partial F}{\partial\bar{p}}\frac{\partial G}{\partial\bar{q}}-\frac{\partial F}{\partial\bar{q}}\frac{\partial G}{\partial\bar{p}},
\end{gather*}
by the Moyal bracket defined as~\cite{moyal1949}
\begin{gather*}
\{F, G\} _{\rm MB} :=\frac{1}{{\rm i}\theta}\left(F \star G-G \star F \right) = \frac{2}{\theta} F\sin\left[ \frac{\theta}{2} \left( \frac{\overleftarrow{\partial}}{\partial\bar{p}}\frac{\overrightarrow{\partial}}{\partial\bar{q}} -\frac{\overleftarrow{\partial}}{\partial\bar{q}}\frac{\overrightarrow{\partial}}{\partial\bar{p}} \right) \right] G,
\end{gather*}
with the $\star$-product defined as
\begin{gather*}
f \star g: = f \exp\left[\frac{{\rm i}\theta}{2} \left( \frac{\overleftarrow{\partial}}{\partial\bar{p}}\frac{\overrightarrow{\partial}}{\partial\bar{q}} -\frac{\overleftarrow{\partial}}{\partial\bar{q}}\frac{\overrightarrow{\partial}}{\partial\bar{p}} \right) \right] g.
%\label{eq:The Moyal *-product}
\end{gather*}
Thus the Moyal algebra is a deformation of the Poisson algebra. Using the well-known Taylor expansion of the sine function we obtain
\begin{gather*}
\{F, G\} _{\rm MB}= \sum_s^\infty -2{\rm i}{\theta}^{2s}\big(F *^{2s+1} G\big),%\label{eq: Taylor expansion of the sine}
\end{gather*}
where the product $*^{2s+1}$ has the generic form
\begin{gather*}
f(x)*^{r}g(x)=\dfrac{1}{r!}\left(\frac{{\rm i}}{2}\right)^{r}\varepsilon^{\mu_{1}\nu_{1}} \cdots\varepsilon^{\mu_{r}\nu_{r}}\partial_{\mu_{1}}\cdots\partial_{\mu_{r}}f(x) \partial_{\nu_{1}}\cdots\partial_{\nu_{r}}g(x).%\label{eq:Moyal *r-order}
\end{gather*}
The above formulation can be also implemented by considering a noncommutative matrix
\begin{gather}
\theta^{\mu\nu} :=\theta\varepsilon^{\mu\nu}=\theta\delta_{\bar{\imath}}^{\mu}\delta_{\bar{\jmath}}^{\nu}\epsilon^{\bar{\imath}\bar{\jmath}} = \theta\left(\begin{matrix}
0 & 0 & 0 & 0\\
0 & 0 & 0 & 0\\
0 & 0 & 0 & 1\\
0 & 0 & -1 & 0
\end{matrix}\right).\label{eq:NC parameter}
\end{gather}
with the $\star$-product defined as
\begin{gather}
f \star g: = f \exp\left( \frac{{\rm i}}{2} \theta^{\mu\nu} \frac{\overleftarrow{\partial}}{\partial x^\mu }\frac{\overrightarrow{\partial}}{\partial x^\nu} \right) g.
\label{starproduct}
\end{gather}
In the above, $\varepsilon^{\mu\nu} = - \varepsilon^{\nu\mu}$ is an anti-symmetric object, and $\epsilon^{\bar{\imath}\bar{\jmath}}$ is the standard Levi-Civita tensor in two dimensions. Following~\cite{STRACHAN199263} and~\cite{Plebanski:1995gk}, we briefly review the iterative method for constructing a set of differential equations for the Moyal deformation of the first heavenly equation. For this purpose, we consider then a series development in powers of $\theta$ for $\widehat{\Omega}$
\begin{gather}
\widehat{\Omega} = \sum_{n=0}^\infty \theta^{n} \Omega^{(n)} , \qquad n=0,1,\ldots.
\label{eq:deformed potential}
\end{gather}
Here $\Omega^{(n)}$ are functions to be determined and $\Omega^{(0)} =: \Omega$ is the classical K\"ahler potential. Plugging this expansion into the integrable deformation of Pleba\'nski's (first heavenly) equation
\begin{gather*}
\big\{\widehat{\Omega}_{p}, \widehat{\Omega}_{q}\big\}_{\rm MB}=1,
%\label{eq:integrable deformation of Plebanski=002018s first heavenly equation}
\end{gather*}
we obtain the expression
\begin{gather}
\big\{ \widehat \Omega_{p}, \widehat\Omega_{q}\big\} _{\rm MB} =-2{\rm i} \sum_{r=0}^\infty \theta^{r} \sum_{s=0}^{\left[\frac{r}{2}\right]} \sum_{m=0}^{r-2s} \big(\partial_{p}\Omega^{(m)} *^{2s+1} \partial_{q}\Omega^{(r-m-2s)}\big)=1.
\label{eq:deformation of the first heavenly equation}
\end{gather}
We compare now the coefficients of the same powers of $\theta$ in both sides of this equation; for $r=0$ we find the first heavenly equation~(\ref{eq:Monge-Amper=0000E9 equation-1-1}). For any $r\geq1$, equation~(\ref{eq:deformation of the first heavenly equation}) gives the condition
\begin{gather}
\sum_{s=0}^{\left[\frac{r}{2}\right]} \sum_{m=0}^{r-2s} \big(\partial_{p}\Omega^{(m)} *^{2s+1} \partial_{q}\Omega^{(r-m-2s)}\big)=0.
\label{eq:r-st order of first heavenly equation}
\end{gather}
To analyse the lowest order modifications to the first heavenly equation, we set $r=1$ to obtain
\begin{gather}
\Omega_{p\bar{p}}\Omega_{q\bar{q}}^{(1)}+\Omega_{p\bar{p}}^{(1)}\Omega_{q\bar{q}}-\Omega_{p\bar{q}}\Omega_{q\bar{p}}^{(1)}-\Omega_{p\bar{q}}^{(1)}\Omega_{q\bar{p}}=0.
\label{eq:first heavenly equation}
\end{gather}
Therefore, once $\Omega$ is known, equation~(\ref{eq:first heavenly equation}) becomes a linear partial differential equation for the first order corrections $\Omega^{(1)}$. The deformed potential $\widehat \Omega$ defined as before is not necessarily K\"ahler-like, we need to impose additional conditions to guarantee that it will be; these conditions will be discussed in the next section.

\section{Integrable systems \label{sec:Integrable-Systems}}

Several multidimensional integrable systems were discussed in~\cite{Strachan:1996gx} assuming a symplectic manifold with some associated $\star$-product. All these systems share the characteristic feature that they have associated a 2-form $\widehat{\boldsymbol{\Omega}}$ which satisfies the equations
\begin{gather}
{\rm d}\widehat{\boldsymbol{\Omega}} = 0,\label{eq:integrability conditions1}\\
\widehat{\boldsymbol{\Omega}} \wedge \widehat{\boldsymbol{\Omega}} = 0.\label{eq:integrability conditions2}
\end{gather}
These equations contain the integrability conditions of the systems in a concise geometric way. In the following, we focus on the implications of these relations for the particular case of an integrable deformation of self-dual vacuum Einstein equations.

\subsection{The self-dual vacuum Einstein equations} \label{subsec:The-Self-Dual-Vacuum Einstein Equations}

Consider a real manifold $(\mathcal{M}, g)$ of dimension four and metric
\begin{gather*}
{\rm d}s^{2}=g_{ab}{\rm d}\phi^{a}{\rm d}\phi^{b}, \qquad a=1,\dots, 4.
\end{gather*}
We assume that the associated Levi-Civita covariant derivative is torsionless, i.e., $D_{k}g_{ab}=0$. Locally it is possible to introduce complex
coordinates $\phi^{a}=(\tau^{i},\tau^{\bar{\imath}})$, $i, \bar{\imath}=1, 2$, such that the metric reads
\begin{gather}
{\rm d}s^{2}=g_{i\bar{\jmath}}{\rm d}\tau^{i}{\rm d}\tau^{\bar{\jmath}}+g_{\bar{\imath}j}{\rm d}\tau^{\bar{\imath}}{\rm d}\tau^{j}=2g_{i\bar{\jmath}}{\rm d}\tau^{i}{\rm d}\tau^{\bar{\jmath}}.\label{eq:K=0000E4hler manifold}
\end{gather}
Note that this metric is real since it was so in the original coordinates; as a consequence we have
\begin{gather*}
g_{i\bar{\jmath}}=g_{j\bar{\imath}}^{*}, \qquad g_{ij}=g_{\bar{\jmath}\bar{\imath}}^{*}=0.
\end{gather*}
The indexes $i$ and $\bar{\imath}$ are called holomorphic and antiholomorphic respectively; the standard convention is to write the holomorphic index first.

Complex 4-metrics of Euclidean signature with vanishing Ricci tensor and anti-self-dual Weyl tensor correspond to anti-self-dual vacuum Einstein solutions. Using the fact that these metrics are K\"ahler, they may be written in terms of the K\"ahler potential $\Omega$ as $ g_{i\bar{\jmath}} := \Omega_{i\bar{\jmath}}:= \partial_{i\bar{\jmath}}\Omega$. It follows that
\begin{gather}
g_{\mu\nu}=\big(\delta_{\mu}^{i}\delta_{\nu}^{\bar{\jmath}}+\delta_{\nu}^{i}\delta_{\mu}^{\bar{\jmath}}\big)\Omega_{i\bar{\jmath}} =\left(\begin{matrix}
0 & \Omega_{i\bar{\jmath}}\\
\Omega_{j\bar{\imath}} & 0
\end{matrix}\right). \label{eq:metric in complex coordinates}
\end{gather}

As mentioned before, a noncommutative deformation of the integrability conditions equa\-tions~(\ref{eq:integrability conditions1}) and~(\ref{eq:integrability conditions2}) was proposed in~\cite{Strachan:1996gx}. More specifically, it was noted that if $\widehat{\boldsymbol{\Omega}}$ is the deformed 2-form
\begin{gather}
\widehat{\boldsymbol{\Omega}}={\rm d}p\wedge {\rm d}q+\lambda\big(\widehat{\Omega}_{p\bar{p}}{\rm d}p\wedge {\rm d}\bar{p}+\widehat{\Omega}_{p\bar{q}}{\rm d}p\wedge {\rm d}\bar{q}+\widehat{\Omega}_{q\bar{p}}{\rm d}q\wedge {\rm d}\bar{p}+\widehat{\Omega}_{q\bar{q}}{\rm d}q\wedge {\rm d}\bar{q}\big)+\lambda^{2}{\rm d}\bar{p}\wedge {\rm d}\bar{q},
\label{eq:the deformed 2-form}
\end{gather}
then it clearly satisfies the condition ${\rm d}\widehat{\boldsymbol{\Omega}}=0$. Furthermore, it is straightforward to see that
\begin{gather*}
\widehat{\boldsymbol{\Omega}} \wedge \widehat{\boldsymbol{\Omega}} = \lambda^{2}\big(\big\{\widehat{\Omega}_{p}, \widehat{\Omega}_{q}\big\}_{\rm MB}-1\big){\rm d}p\wedge {\rm d}\bar{p}\wedge {\rm d}q\wedge {\rm d}\bar{q}.
\end{gather*}
The right hand side of this equation vanishes if $\widehat\Omega$ satisfies the deformed Pleba\'nski equation. This result means that a Moyal deformation of the first heavenly equation preserves its integrability. Furthermore, we have the important result that a perturbative solution equation~(\ref{eq:deformed potential}) exists. This result implies that equation~(\ref{eq:metric in complex coordinates}) generalises to
\begin{gather}
g_{\mu\nu}^{(n)}=\big(\delta_{\mu}^{i}\delta_{\nu}^{\bar{\jmath}}+\delta_{\nu}^{i}\delta_{\mu}^{\bar{\jmath}}\big)\Omega_{i\bar{\jmath}}^{(n)}.\label{eq:deformed metric in complex coordinates}
\end{gather}

\subsection{Deformed properties} \label{subsec:Deformed-properties}

As we previously mentioned, we want to construct a four-dimensional Moyal deformed integrable K\"ahler manifold. For this purpose, we impose that the deformed K\"ahler potential $\widehat{\Omega}$ must share the same features and properties of the undeformed system; they are ($\hat{g}_{i\bar{\jmath}}  := \partial_i \partial_{\bar{\jmath}} \widehat\Omega$)
\begin{enumerate}\itemsep=0pt
\item[1)] the 2-form $\widehat{\boldsymbol{\Omega}} := \widehat{\Omega}_{i\bar{\jmath}} {\rm d}x^i \wedge {\rm d} \bar x^{\bar{\jmath}} := 2 {\rm i} \hat{g}_{i\bar{\jmath}} {\rm d}x^i \wedge {\rm d} \bar x^{\bar{\jmath}}$ should be closed, i.e., ${\rm d}\widehat{\boldsymbol{\Omega}} = 0$,
\item[2)] the metric coefficients $\widehat{\Omega}_{i\bar{\jmath}}$ should be real (hermitian property),
\item[3)] the determinant of $\widehat{\Omega}_{i\bar{\jmath}}$ should be equal to one, i.e., $\det\widehat{\Omega}_{i\bar{\jmath}} := \widehat \Omega_{p\bar p} \star \widehat \Omega_{q\bar q} - \widehat \Omega_{p\bar q} \star \widehat \Omega_{q\bar p} =1$.
\end{enumerate}

\looseness=-1 The first of the above conditions can be analysed by fixing for the 2-form $\widehat{\boldsymbol{\Omega}}$, the same functional form as that of equation~(\ref{eq:the deformed 2-form}). This fact implies that the following condition must hold
\begin{gather*}
\partial_{k}\Omega_{i\bar{\jmath}}^{(n)}=0=\partial_{\bar{k}}\Omega_{i\bar{\jmath}}^{(n)}, \qquad k = p, q, \qquad \bar k = \bar p, \bar q.
\end{gather*}
We now write equation~(\ref{eq:r-st order of first heavenly equation}) as
\begin{gather*}
\sum_{m=0}^r \big(\partial_{p}\Omega^{(m)} *^{1} \partial_{q}\Omega^{(r-m)}\big) + \sum_{s=1}^{\left[\frac{r}{2}\right]} \sum_{m=0}^{r-2s} \big(\partial_{p}\Omega^{(m)} *^{2s+1} \partial_{q}\Omega^{(r-m-2s)}\big) = 0.
\end{gather*}
The second sum vanishes because of the first property imposed on $\Omega^{(n)}$; we have thus
\begin{gather}
\sum_{m=0}^r \big(\partial_{p}\Omega^{(m)} *^{1} \partial_{q}\Omega^{(r-m)}\big)=\sum_{m=0}^r \epsilon^{\bar{\imath}\bar{\jmath}}\Omega_{p\bar{\imath}}^{(m)}\Omega_{q\bar{\jmath}}^{(r-m)}=0.
\label{eq:simplification of the deformed first heavenly eq}
\end{gather}
Therefore, we can write equation~(\ref{eq:deformation of the first heavenly equation}) in terms of the Poisson bracket as
\begin{gather*}
\big\{\widehat{\Omega}_{p}, \widehat{\Omega}_{q}\big\}_{\rm PB}=1.
\end{gather*}
The second condition implies that the corresponding metric coefficients are hermitian as in the commutative case; therefore, each perturbation $\Omega^{(n)}$ is also hermitian. We have then
\begin{gather*}
\Omega_{i\bar{\jmath}}^{(n)}=\Omega_{j\bar{\imath}}^{\dagger(n)}, \qquad \Omega_{ij}^{(n)}=\Omega_{\bar{\imath}\bar{\jmath}}^{\dagger(n)}=0,
\end{gather*}
where the last two conditions hold because we have a deformed K\"ahler manifold.

The third property can be imposed from the curvature condition equation~(\ref{eq:Monge-Amper=0000E9 equation-1-1}) since the deformed first heavenly equation admits a rewriting as a simple determinant. We know that in the case of the non-deformed K\"ahler manifold, the determinant of the metric tensor is equal to one. In the case of the deformed case the deformed metric tensor $\widehat{\Omega}_{i\bar{\jmath}}$ should then satisfy the same property. We demand that by definition
\begin{gather*}
\det\widehat{\Omega}_{i\bar{\jmath}} := \widehat{\Omega}_{p\bar{p}}\widehat{\Omega}_{q\bar{q}} - \widehat{\Omega}_{p\bar{q}}\widehat{\Omega}_{q\bar{p}}=1.
\end{gather*}
The above equation is equivalent to $\det\widehat{\Omega}_{i\bar{\jmath}} = \epsilon^{\bar{k} \bar{l}}\widehat{\Omega}_{p\bar{k}}\widehat{\Omega}_{q\bar{l}} = 1$. Using the power series expansion in equation~(\ref{eq:deformed potential}) for $\widehat{\Omega}$, we get the different contributions to $\det\widehat{\Omega}_{i\bar{\jmath}}$ order by order on $\theta$
\begin{gather*}
\det\widehat{\Omega}_{i\bar{\jmath}} = \{\Omega_{p}, \Omega_{q}\}_{\rm PB} +  \sum_{r=0}^\infty \sum_{\substack{m,n=0\\ m+n = r\ge 1}}^r \epsilon^{\bar k \bar l} \theta^{m+n}\Omega_{p\bar k}^{(m)}\Omega_{q\bar l}^{(n)}.
\end{gather*}
The first term in this expression is the Moyal deformation of the first heavenly equation; its value is equal to one. Therefore, we conclude that the
second term should vanish, namely
\begin{gather*}
\sum_{\substack{m,n=0\\ m+n = r\geq 1}}^r\epsilon^{\bar k\bar l} \theta^{m+n}\Omega_{p\bar k}^{(m)}\Omega_{q\bar l}^{(n)}=0. \label{eq:residues}
\end{gather*}
In this equation are encoded all the combinations of order $\theta^j$ such that $j > r$.

\section{Noncommutative gravity}\label{sec:Noncommutative-gravity}

In the previous section, we studied the noncommutative deformation of the K\"ahler potential using the Moyal deformation of the first heavenly equation. The deformation functions $\Omega^{(n)}$ are unknown; each one of them satisfy their respective differential equations obtained from equation~(\ref{eq:r-st order of first heavenly equation}). In this section, we give an ansatz for the deformed K\"ahler potentials $\Omega^{(n)}$ appearing in equation~(\ref{eq:simplification of the deformed first heavenly eq}). As we will see later, the deformed functions $\Omega^{(n)}$ can be expressed in terms of a deformed local frame or vierbein $\hat{e}_{\mu}^{a}(x, \theta)$; after an integration procedure, the deformed K\"ahler potential $\widehat\Omega$ will be written in terms of the vierbein.

\subsection{Deformed gauge fields\label{subsec:Deformed-gauge-fields}}

For some time now, a subject of interest has been the construction of consistent noncommutative deformations of Einstein gravity. Following the
standard procedure to construct noncommutative gauge and scalar field theories~\cite{Douglas:2001ba,Szabo:2001kg}, noncommutative versions of the Einstein-Hilbert action have been obtained by replacing the ordinary product by the noncommutative Moyal product in equation~(\ref{starproduct}). The noncommutative structure of spacetime is then
\begin{gather*}
\big[x^{\mu}, x^{\nu}\big]_\star = {\rm i}\theta^{\mu\nu},
\end{gather*}
where the elements $\theta^{\mu\nu} $ are constant (canonical) parameters and antisymmetric, i.e., $\theta^{\mu\nu}=-\theta^{\nu\mu}$.

In this approach, we can introduce a noncommutative metric as
\begin{gather}
\hat{g}_{\mu\nu} = \frac{1}{2}\big(\hat{e}_{\mu}^{a} \star \hat{e}_{\nu}^{b}+\hat{e}_{\nu}^{a} \star \hat{e}_{\mu}^{b}\big)\eta_{ab},
\label{eq: Def: Deformed Metric}
\end{gather}
in terms of a vierbein $\hat{e}_{\mu}^{a}(x, \theta)$ and the Minkowski metric $\eta_{ab}$; the vierbein $\hat{e}_{\mu}^{a}(x, \theta)$ reduces to the commutative one when $\theta = 0$. The metric $\hat{g}_{\mu\nu}$ is symmetric by construction and real even if the deformed tetrad fields $\hat{e}_{\mu}^{a}(x, \theta)$ are complex quantities. For $\theta=0$, we identify this metric with the commutative metric field
\begin{gather*}
 \hat{g}_{\mu\nu}(x, \theta)\big|_{\theta=0} = g_{\mu\nu} = e_{\mu}^{a}e_{\nu}^{b}\eta_{ab}.
%\label{eq:G vierbein}
\end{gather*}
We want now to construct a Moyal deformed spacetime with associated deformed vierbein and metric, sharing the same properties of the undeformed spacetime, and such that the deformed K\"ahler potential discussed in the Section~\ref{subsec:Deformed-properties} exists. For this purpose, we first introduce the vector fields $\hat{e}_{\mu}^{b}$ as
\begin{gather*}
\hat{e}_{\mu}^{b} = e_{\mu}^{b}+\theta^{k\lambda}e_{\mu k\lambda}^{b}+\cdots+\theta^{k_{1}\lambda_{1}}\cdots\theta^{k_{n}\lambda_{n}}e_{\mu k_{1}\lambda_{1}\cdots k_{n}\lambda_{n}}^{b} + \cdots,
\end{gather*}
where the elements $e_{\mu k_{1}\lambda_{1}\cdots k_{n}\lambda_{n}}^{b}$ are to be found. With this series expansion in powers of $\theta$, we write then the metric tensor as
\begin{gather}
\hat{g}_{\mu\nu} = g_{\mu\nu}+{\rm i}\theta^{k\lambda}g_{\mu\nu k\lambda}^{(1)}+\mathcal{O}\big(\theta^{2}\big),\label{eq:Metric Deformed}
\end{gather}
up to first order on $\theta$.

Since we are interested in making compatible the noncommutative deformation of the K\"ahler metric equation~(\ref{eq:deformed metric in complex coordinates}), with the deformation in equation~(\ref{eq:Metric Deformed}), let us assume the following decomposition for the verbein
\begin{gather*}
\theta^{k\lambda}e_{\mu k\lambda}^{b}=\theta^{k\lambda}\mathcal{P}_{k\lambda} e_{\mu}^{(1)b},
\end{gather*}
to first order on $\theta$; here $\mathcal{P}_{k\lambda}$ and $e_{\mu}^{(1)b}$ are unknown quantities. For the $n$-th order we generalize this ansatz to
\begin{gather*}
\theta^{k_{1}\lambda_{1}}\cdots\theta^{k_{n}\lambda_{n}}e_{\mu k_{1}\lambda_{1}\cdots k_{n}\lambda_{n}}^{b}=\big(\theta^{k\lambda}\mathcal{P}_{k\lambda}\big)^{n} e_{\mu}^{(n)b}.
\end{gather*}
We now impose the condition
\begin{gather}
\hat{e}_{\mu}^{b}=\sum_{n=0}^\infty \theta^{n} e_{\mu}^{(n)b}, \qquad e_\mu^{(0) b} := e_\mu^b,
\label{eq:vierbein NC}
\end{gather}
implying $\theta^{k\lambda}\mathcal{P}_{k\lambda}=\theta$; it is easy to show now that if we choose $\mathcal{P}_{k\lambda}:=\partial_{k}\Omega_{p}\partial_{\lambda}\Omega_{q}$ where $\Omega$ is the undeformed K\"ahler potential, then equation~(\ref{eq:vierbein NC}) is satisfied, where $\theta^{k\lambda}$ is given by equation~(\ref{eq:NC parameter}). In a similar way as for the vierbein, the tensor metric equation~(\ref{eq:Metric Deformed}) takes the general form
\begin{gather*}
\hat{g}_{\mu\nu}=\sum_{n=0}^\infty \theta^{n} g_{\mu\nu}^{(n)},
\end{gather*}
where the $g_{\mu\nu}^{(n)}$'s are tensor fields written in terms of the $e_{\mu}^{(n)a}$'s; we fix their form as follows: first, to make compatible this deformation with the structure of a K\"ahler manifold, we need to impose the constraint $\partial_{\alpha}g_{\mu\nu}^{(n)}=0$ with $x^\alpha = (p, q,\bar p, \bar q)$. This condition implies the property $\partial_{\mu}e_{\nu}^{(n)a}=0$ for the vierbein; equation~(\ref{eq: Def: Deformed Metric}) simplifies then to
\begin{gather}
\hat{g}_{\mu\nu} = \hat{e}_{\mu}^{a}\hat{e}_{\nu}^{b}\eta_{ab}.
\label{eq:K=0000E4hler Moyal Metric Deformed}
\end{gather}
Using now the expansion of $\hat{e}_{\mu}^{a}$ in powers of $\theta$ into equation~(\ref{eq:K=0000E4hler Moyal Metric Deformed}), and equating the coefficients of the same power of $\theta$ in both sides of the equation, we obtain the $n$-th tensor field $g_{\mu\nu}^{(n)}$ in terms of the deformed tetrad $e_{\mu}^{(n)a}$ as
\begin{gather*}
g_{\mu\nu}^{(n)} = \sum_{m=0}^n e_{\mu}^{(m)a}e_{\nu}^{(n-m)b}\eta_{ab},\qquad n=0,1,\ldots.
\end{gather*}

\section{The deformed K\"ahler potential}\label{sec:The-Deformed-K=0000E4hler Potential}

According with \cite{Plebanski1975}, self-dual gravity can be parametrised in terms of the complex coordinates $x^{\mu}=\{p, q, \bar{p}, \bar{q}\}$ and the resulting spacetime has the estructure of a K\"ahler manifold. In our case, the K\"ahler spacetime given in equation~(\ref{eq:K=0000E4hler manifold}) can be written in terms of the classical vierbein $e_{\mu}^{a}$ and the local flat spacetime metric $\eta_{ab}$ defined as
\begin{gather}
e_{\mu}^{a} := \left(\begin{matrix}
0 & 0 & \Omega_{p\bar{p}} & \Omega_{p\bar{q}}\\
1 & 0 & 0 & 0\\
0 & 0 & \Omega_{q\bar{p}} & \Omega_{q\bar{q}}\\
0 & 1 & 0 & 0
\end{matrix}\right),
\qquad
\eta_{ab} := \left(\begin{matrix}
0 & 1 & 0 & 0\\
1 & 0 & 0 & 0\\
0 & 0 & 0 & 1\\
0 & 0 & 1 & 0
\end{matrix}\right).\label{eq:non deformed vierbein}
\end{gather}
We now construct an ansatz for the deformed potentials $\Omega^{(n)}$ in terms of the deformed vierbein~$e_{\mu}^{(n)a}$. Since we want that all the properties listed in Section~\ref{subsec:Deformed-properties} hold, we write first
\begin{gather}
g_{\mu\nu}^{(n)} = \sum_{m=0}^n e_{\mu}^{(m)a}e_{\nu}^{(n-m)b}\eta_{ab} = \big(\delta_{\mu}^{i}\delta_{\nu}^{\bar{\jmath}}+\delta_{\nu}^{i}\delta_{\mu}^{\bar{\jmath}}\big)\Omega_{i\bar{\jmath}}^{(n)}.
\label{eq:metric (n-th order)}
\end{gather}
The simplest ansatz for the vierbein $e_{\mu}^{(n)a}$ that satisfies $g_{ij}^{(n)}=g_{\bar{\jmath}\bar{\imath}}^{(n)*}=0$, is
\begin{gather*}
e_{\mu}^{(n)a}=\left(\begin{matrix}
0 & 0 & e_{\bar{p}}^{(n)1} & e_{\bar{q}}^{(n)1}\\
e_{p}^{(n)2} & e_{q}^{(n)2} & 0 & 0\\
0 & 0 & e_{\bar{p}}^{(n)3} & e_{\bar{q}}^{(n)3}\\
e_{p}^{(n)4} & e_{q}^{(n)4} & 0 & 0
\end{matrix}\right).
\end{gather*}
As we discussed previously, the vierbein $e_{\mu}^{(n)a}$ must have the property $\partial_{\mu}e_{\nu}^{(n)a}=0$, $\mu = p, q, \bar p, \bar q$. Therefore, we shall assume the following dependence
\begin{gather}
e_{\mu}^{(n+1)a}= e_{\mu}^{(n+1)a} \big( \Omega_{p\bar{p}}^{(n)}, \Omega_{q\bar{q}}^{(n)}, \Omega_{p\bar{q}}^{(n)}, \Omega_{q\bar{p}}^{(n)} \big).
\label{eq: vierbein functional dependence}
\end{gather}

The deformed K\"ahler potential $\Omega^{(n+1)}$ to order $n+1$ will depend explicitly on the vierbein to order $n+1$ and $n$, i.e., on $e_{\mu}^{(n+1)a}$ and $e_{\mu}^{(n)a}$ respectively. We begin by analising the deformation to first order: from equation~(\ref{eq:metric (n-th order)}) we obtain
\begin{gather*}
\Omega_{i\bar{\jmath}}^{(1)}=e_{(i}^{(1)a}e_{\bar{\jmath})}^{b}\eta_{ab}.
\end{gather*}
Using the undeformed vierbein $e_{\mu}^{a}$ given in equation~(\ref{eq:non deformed vierbein}), we write explicitly
\begin{gather}
\Omega_{q\bar{q}}^{(1)} = e_{\bar{q}}^{(1)3}+e_{q}^{(1)2}\Omega_{p\bar{q}}+e_{q}^{(1)4}\Omega_{q\bar{q}},
\label{eq:Gqq first order}\\
\Omega_{q\bar{p}}^{(1)} = e_{\bar{p}}^{(1)3}+e_{q}^{(1)2}\Omega_{p\bar{p}}+e_{q}^{(1)4}\Omega_{q\bar{p}},
\label{eq:Gqp first order}\\
\Omega_{p\bar{p}}^{(1)} = e_{\bar{p}}^{(1)1}+e_{p}^{(1)2}\Omega_{p\bar{p}}+e_{p}^{(1)4}\Omega_{q\bar{p}},
\label{eq:Gpp first order}\\
\Omega_{p\bar{q}}^{(1)} = e_{\bar{q}}^{(1)1}+e_{p}^{(1)2}\Omega_{p\bar{q}}+e_{p}^{(1)4}\Omega_{q\bar{q}}.
\label{eq:Gpqfirst order}
\end{gather}
To obtain the K\"ahler potential up to first order, we need to solve equations~(\ref{eq:Gqq first order})--(\ref{eq:Gpqfirst order}).
If we integrate first the above equations with respect to the anti-holomorphic variables $x^{\bar{\imath}} = \{\bar{p}, \bar{q}\}$, we need to calculate a set of integrals of the form $\int e_{\alpha}^{(1)A}\Omega_{\beta\mu} {\rm d}x^{\nu}$. After an integration by parts, we see that
\begin{gather*}
\int e_{\alpha}^{(1)A}\Omega_{\beta\mu} {\rm d}x^{\nu}=\begin{cases}
e_{\alpha}^{(1)A}\Omega_{\beta\mu} x^{\nu}+C_{i}^{(1)}, & \text{if} \ \mu\neq\nu,\\
e_{\alpha}^{(1)A}\Omega_{\beta}+C_{i}^{(1)}, & \text{if} \ \mu=\nu,
\end{cases}
\end{gather*}
where we used the properties $\partial_{\mu}e_{\nu}^{(1)a}=0$, $\partial_{\alpha}\Omega_{\mu\nu}=0$ of the vierbein and the K\"ahler potential respectively. In consequence, we have
\begin{gather}
\Omega_{q}^{(1)} = e_{\bar{q}}^{(1)3}\bar{q}+e_{q}^{(1)2}\Omega_{p}+e_{q}^{(1)4}\Omega_{q}+C_{1}^{(1)}(p, q, \bar{p}),
\label{eq:Gqq first order-1}\\
\Omega_{q}^{(1)} = e_{\bar{p}}^{(1)3}\bar{p}+e_{q}^{(1)2}\Omega_{p}+e_{q}^{(1)4}\Omega_{q}+C_{2}^{(1)}(p, q, \bar{q}),
\label{eq:Gqp first order-1}\\
\Omega_{p}^{(1)} = e_{\bar{p}}^{(1)1}\bar{p}+e_{p}^{(1)2}\Omega_{p}+e_{p}^{(1)4}\Omega_{q}+C_{3}^{(1)}(p, q, \bar{q}),
\label{eq:Gpp first order-1}\\
\Omega_{p}^{(1)} = e_{\bar{q}}^{(1)1}\bar{q}+e_{p}^{(1)2}\Omega_{p}+e_{p}^{(1)4}\Omega_{q}+C_{4}^{(1)}(p, q, \bar{p}),
\label{eq:Gpqfirst order-1}
\end{gather}
where $C_{1}^{(1)}=C_{1}^{(1)}(p, q, \bar{p})$, $C_{2}^{(1)}=C_{2}^{(1)}(p, q, \bar{q})$, $C_{3}^{(1)}=C_{3}^{(1)}(p, q, \bar{q})$ and $C_{4}^{(1)}=C_{4}^{(1)}(p, q, \bar{p})$ are functions of their arguments. We determine these functions by comparing equations~(\ref{eq:Gqq first order-1})--(\ref{eq:Gqp first order-1}) and (\ref{eq:Gpp first order-1})--(\ref{eq:Gpqfirst order-1}); we conclude that $C_{1}^{(1)}=e_{\bar{p}}^{(1)3}\bar{p}$, $C_{2}^{(1)}=e_{\bar{q}}^{(1)3}\bar{q}$, $C_{3}^{(1)}=e_{\bar{q}}^{(1)1}\bar{q}$ and $C_{4}^{(1)}=e_{\bar{p}}^{(1)1}\bar{p}$. Therefore, we obtain the following two expressions
\begin{gather}
\Omega_{q}^{(1)} = e_{\bar{p}}^{(1)3}\bar{p}+e_{\bar{q}}^{(1)3}\bar{q}+e_{q}^{(1)2}\Omega_{p}+e_{q}^{(1)4}\Omega_{q},\label{eq:Gqq first order-1-1}\\
\Omega_{p}^{(1)} = e_{\bar{p}}^{(1)1}\bar{p}+e_{\bar{q}}^{(1)1}\bar{q}+e_{p}^{(1)2}\Omega_{p}+e_{p}^{(1)4}\Omega_{q}.\label{eq:Gpp first order-1-1}
\end{gather}
Following the same procedure as before, after an integration by parts of equations~(\ref{eq:Gqq first order-1-1}) and~(\ref{eq:Gpp first order-1-1}) with respect to the holomorphic variables $q$ and $p$ respectively, the K\"ahler potential takes the unique form
\begin{gather}
\Omega^{(1)} = \big(e_{\bar{p}}^{(1)1}p+e_{\bar{p}}^{(1)3}q\big)\bar{p}+\big(e_{\bar{q}}^{(1)1}p+e_{\bar{q}}^{(1)3}q\big)\bar{q}+e_{q}^{(1)2}\Omega_{p}q+e_{p}^{(1)4}\Omega_{q}p\nonumber\\
\hphantom{\Omega^{(1)} =}{} + \big( e_{p}^{(1)2} + e_{q}^{(1)4} \big) \Omega.\label{kpneq1}
\end{gather}
In general, a straightforward calculation shows that the K\"ahler potential up to the $n$-th order has the expression
\begin{gather*}
\Omega^{(n)} = \big(e_{\bar{p}}^{(n)1}p+e_{\bar{p}}^{(n)3}q\big)\bar{p}+\big(e_{\bar{q}}^{(n)1}p+e_{\bar{q}}^{(n)3}q\big)\bar{q}
+e_{q}^{(n)2}\Omega_{p}q+e_{p}^{(n)4}\Omega_{q}p+\big(e_{p}^{(n)2}+e_{q}^{(n)4}\big)\Omega
\\
\hphantom{\Omega^{(n)} =}{} +\sum_{m=1}^{n-1}\big(e_{p}^{(m)a}p+e_{q}^{(m)a}q\big)\big(e_{\bar{p}}^{(n-m)b}\bar{p}+e_{\bar{q}}^{(n-m)b}\bar{q}\big)\eta_{ab}.
\end{gather*}
When $n=1$, we recover equation~(\ref{kpneq1}).

\subsection{Solutions for the K\"ahler potential to first order}\label{subsec:Solutions-for-the K=0000E4hler potential to first order}

As a particular example of the previous approach, we consider now in detail the deformation of the K\"ahler potential and the vierbein up to first order on the noncommutative parameter $\theta$. We recall that the curvature condition can be formulated as a simple determinant, that is
\begin{gather*}
\det\widehat{\Omega}_{i\bar{\jmath}}=\widehat{\Omega}_{p\bar{p}}\widehat{\Omega}_{q\bar{q}}-\widehat{\Omega}_{p\bar{q}}\widehat{\Omega}_{q\bar{p}}=1,
\end{gather*}
where $\widehat{\Omega}=\Omega+\theta\Omega^{(1)}$ up to first order. Substituting this expression in the determinant condition, we obtain the following equations
\begin{gather}
\Omega_{p\bar{p}} \Omega_{q\bar{q}} - \Omega_{p\bar{q}} \Omega_{q\bar{p}} = 1,\label{eq:M-A zero order}\\
\Omega_{p\bar{p}} \Omega_{q\bar{q}}^{(1)}+\Omega_{p\bar{p}}^{(1)}\Omega_{q\bar{q}} - \Omega_{p\bar{q}}\Omega_{q\bar{p}}^{(1)}-\Omega_{p\bar{q}}^{(1)}\Omega_{q\bar{p}} = 0.
\label{eq:M-A first order}
\end{gather}
Equations~(\ref{eq:M-A zero order}) and (\ref{eq:M-A first order}) are the Monge--Amp\`ere equations to zero and first order respectively.
If we substitute now equations~(\ref{eq:Gqq first order})--(\ref{eq:Gpqfirst order}) into the heavenly equation to first order equation~(\ref{eq:M-A first order}), we obtain
\begin{gather*}
e_{q}^{(1)4}+e_{p}^{(1)2}+e_{\bar{q}}^{(1)3}\Omega_{p\bar{p}}+e_{\bar{p}}^{(1)1}\Omega_{q\bar{q}}-e_{\bar{p}}^{(1)3}\Omega_{p\bar{q}}-e_{\bar{q}}^{(1)1}\Omega_{q\bar{p}}=0.
\end{gather*}
Therefore, we need to find a form for the vierbein $e_{\mu}^{(1)a}$ such that the previous equation holds. We consider the following two possibilities
\begin{gather*}
\Omega_{p\bar{p}}^{(1)}=C \Omega_{q\bar{p}}^{(1)},\qquad \Omega_{p\bar{q}}^{(1)}=C \Omega_{q\bar{q}}^{(1)},
\end{gather*}
for case I, and
\begin{gather*}
\Omega_{p\bar{p}}^{(1)}=C \Omega_{p\bar{q}}^{(1)},\qquad \Omega_{q\bar{p}}^{(1)}=C \Omega_{q\bar{q}}^{(1)},
\end{gather*}
for case II. We use now equations~(\ref{eq:Gqq first order})--(\ref{eq:Gpqfirst order}) into the previous formulas to obtain, after some simplifications, the following relations among the components of the vierbein to first order
\begin{gather*}
Ce_{q}^{(1)4}=e_{p}^{(1)4}, \qquad Ce_{q}^{(1)2}=e_{p}^{(1)2},\qquad
Ce_{\bar{q}}^{(1)3}=e_{\bar{q}}^{(1)1}, \qquad Ce_{\bar{p}}^{(1)3}=e_{\bar{p}}^{(1)1},
\end{gather*}
for case I, and
\begin{gather*}
e_{\bar{p}}^{(1)1}=Ce_{\bar{q}}^{(1)1}, \qquad e_{\bar{p}}^{(1)3}=Ce_{\bar{q}}^{(1)3},\qquad
e_{p}^{(1)4}=\frac{A}{A'}e_{q}^{(1)4} = A(\Omega_{q\bar{p}}-C\Omega_{q\bar{q}}), \\
e_{p}^{(1)2}=\frac{A}{A'}e_{q}^{(1)2} = A(\Omega_{p\bar{p}}-C\Omega_{p\bar{q}}),
\end{gather*}
for case II. $A$, $A^\prime$ and $C$ are arbitrary constants in the above expressions. After the respective simplifications, the Monge--Amp\`ere equation becomes
\begin{gather}
e_{q}^{(1)4} + e_{p}^{(1)2} + e_{\bar{q}}^{(1)3}(\Omega_{p\bar{p}} - C\Omega_{q\bar{p}}) + e_{\bar{p}}^{(1)1}(C\Omega_{q\bar{q}}-\Omega_{p\bar{q}})=0,
\label{eq:condition Case I}
\end{gather}
for case I and
\begin{gather}
e_{q}^{(1)4} + e_{p}^{(1)2} + e_{\bar{q}}^{(1)3}(\Omega_{p\bar{p}} - C\Omega_{p\bar{q}}) + e_{\bar{p}}^{(1)1}(C\Omega_{q\bar{q}}-\Omega_{q\bar{p}})=0,
\label{eq:condition Case II}
\end{gather}
for case II.

To further proceed, we recall that according to equation~(\ref{eq: vierbein functional dependence}) the vierbein $e_{\mu}^{(1)a}$ should have the functional dependence
\begin{gather*}
e_{\mu}^{(1)a}=e_{\mu}^{(1)a}( \Omega_{p\bar{p}}, \Omega_{q\bar{q}}, \Omega_{p\bar{q}}, \Omega_{q\bar{p}}),
\end{gather*}
where $\Omega_{i\bar{\jmath}}$, with $\{i ,\bar{\jmath}\} = \{p, q, \bar p, \bar q\}$, are the metric coefficients of the undeformed metric tensor. For case I, we choose the following ansatz for the vierbein
\begin{gather*}
e_{q}^{(1)4} = \alpha+\beta\Omega_{p\bar{p}}+\gamma\Omega_{q\bar{q}}+\delta\Omega_{p\bar{q}}+\sigma\Omega_{q\bar{p}},\nonumber \\
e_{p}^{(1)2} = \alpha'+\beta'\Omega_{p\bar{p}}+\gamma'\Omega_{q\bar{q}}+\delta'\Omega_{p\bar{q}}+\sigma'\Omega_{q\bar{p}},\nonumber \\
e_{\bar{q}}^{(1)3} = \alpha''+\beta'' \Omega_{p\bar{p}}+\gamma''\Omega_{q\bar{q}}+\delta''\Omega_{p\bar{q}}+\sigma''\Omega_{q\bar{p}},\nonumber \\
e_{\bar{p}}^{(1)3} = \alpha'''+\beta'''\Omega_{p\bar{p}}+\gamma'''\Omega_{q\bar{q}}+\delta'''\Omega_{p\bar{q}}+\sigma'''\Omega_{q\bar{p}},
\end{gather*}
where $\alpha, \alpha^\prime, \beta, \beta^\prime, \dots$ are arbitrary constants. If we substitute the above equations into equation~(\ref{eq:condition Case I}), we obtain the following relationship between the coefficients
\begin{gather*}
\alpha+\alpha'+\gamma''+C\beta''' = 0,
\qquad
\beta+\beta'+\alpha'' = 0,
\qquad
\gamma+\gamma'+C\alpha''' = 0,
\nonumber \\
\delta+\delta'-\alpha''' = 0,
\qquad
\sigma+\sigma'-C\alpha'' = 0.
\end{gather*}
The solution to this system of coupled linear equations is
\begin{gather*}
 \gamma''= \sigma''' = - \delta'' = -\beta''', \qquad \gamma'''=0=\delta''', \qquad \sigma''=0=\beta'',
\nonumber \\
 C = -\displaystyle\frac{\alpha+\alpha^\prime+\gamma+\gamma'+\gamma^{\prime\prime}+\sigma+\sigma^\prime}{\beta'''-\alpha''+\alpha'''}.
\end{gather*}

For case II, we choose the following ansatz for the vierbein
\begin{gather*}
 e_{\bar{q}}^{(1)1} = \alpha+\beta\Omega_{p\bar{p}}+\gamma\Omega_{q\bar{q}}+\delta\Omega_{p\bar{q}}+\sigma\Omega_{q\bar{p}},
\nonumber \\
 e_{\bar{q}}^{(1)3} = \alpha'+\beta'\Omega_{p\bar{p}}+\gamma'\Omega_{q\bar{q}}+\delta'\Omega_{p\bar{q}}+\sigma'\Omega_{q\bar{p}},
\end{gather*}
where $\alpha, \alpha^\prime, \beta, \beta^\prime, \dots$ are arbitrary constants. Substitution of these equations into equation~(\ref{eq:condition Case II}) leads to
\begin{gather*}
 \gamma'+C\beta = 0, \qquad \alpha'+A' = 0, \qquad \alpha-A = 0,\\
 \delta-\gamma' = 0, \qquad \sigma' - \beta = 0, \qquad \delta+C\sigma' = 0,
\end{gather*}
with solution
\begin{gather*}
\gamma=0=\sigma, \qquad \delta'=0=\beta', \qquad C=-\frac{\gamma'}{\sigma'}=-\frac{\delta}{\beta}.
\end{gather*}
The previous expressions determine the K\"ahler potential that is compatible with the Moyal deformation of the first heavenly equation. We have thus arrived to a multi-parameter family of solutions for the Moyal-deformed K\"ahler potential.

\section[Deformed $\eta$-self dual Riemman metric]{Deformed $\boldsymbol{\eta}$-self dual Riemman metric}\label{sec:Deformed--self-dual Riemman 4- metric}

Following~\cite{C.P.BoyerandJ.D.FinleyIIIdoi:10.1063/1.525479}, we consider the algebraic description of four dimensional non-compact hyper-K\"ahler manifolds that possess at least one abelian isometry, i.e., a translational or rotational symmetry. For this purpose, we start by recalling the rotational character of the corresponding Killing vector fields. By its own definition, a Killing vector field $\xi_{\mu}$ satisfies $\nabla_{(\nu}\xi_{\mu)}=0$, while the self-duality of the anti-symmetric part $\nabla_{[\nu}\xi_{\mu]}$ provides the critical distinction between these two types of Killing vectors field~\cite{C.P.BoyerandJ.D.FinleyIIIdoi:10.1063/1.525479, Gegenberg1984}: $\xi_{\mu}$ is translational if it satisfies the condition{\samepage
\begin{gather*}
\xi_{\alpha;\beta}=\frac{1}{2}\eta\epsilon_{\alpha\beta}^{\quad\mu\nu}\xi_{\mu;\nu}, \qquad \mbox{with} \quad \eta =\pm 1.
%\label{eq:self dual Killing vector field}
\end{gather*}
Otherwise, we say that $\xi_{\mu}$ is rotational.}

On the other hand, let $g_{\alpha\beta}$ be a $\eta$-self dual Riemman 4-metric with Euclidean signature and let $\xi=\xi^{\alpha}\partial_{\alpha}= \partial/\partial\phi$ be a Killing vector of $g_{\alpha\beta}$. Then, locally we may write
\begin{gather}
{\rm d}s^{2}=\frac{1}{V}\left(d\phi+\omega_{i}{\rm d}x^{i}\right)^{2}+\gamma_{ij}{\rm d}x^{i}dx^{j},\label{eq:self dual Riemman 4- metric}
\end{gather}
with $V$, $\omega_{i}$ and $\gamma_{ij}$ being all independent of $\phi$; Greek indices run from 0 to 3 and Latin indices run from~1 to~3.

In~\cite{Olivier1991}, a set of complex coordinates for the Atiyah--Hitchin (AH) metric was found as an alternative procedure to the twistor formalism~\cite{Bakas:1999wq,Ionas:2007gd}. In the analysis of magnetic monopoles interactions, the moduli space admits a metric formulation in the low-energy limit leading to the AH metric and scattering processes may be analysed in this way; the length element of the AH metric is~\cite{1985AtiyahandN.J.Hitchin}
\begin{gather*}
{\rm d}s^{2}=\beta^{2}\gamma^{2}\delta^{2}\frac{({\rm d}k)^{2}}{\big(4k^{2}k'^{2}K^{2}\big)^{2}}+\beta^{2}\sigma_{x}^{2}+\gamma^{2}\sigma_{y}^{2}+\delta^{2}\sigma_{z}^{2},
\end{gather*}
where
\begin{gather*}
\beta\gamma = -K^{2}\big(k'^{2}+u\big),\qquad \gamma\delta = K^{2}\big(k'^{2}-u\big),\qquad \beta\delta = -K^{2}u,
\end{gather*}
and
\begin{gather*}
u = \frac{G(k)}{K(k)}, \qquad G(k) = E(k)-k'^{2}K(k), \qquad k'^{2} = 1-k^{2}.
\end{gather*}
The differential 1-forms $\sigma_x$, $\sigma_y$ and $\sigma_z$ in the metric are invariant under ${\rm SU}(2)$~\cite{1985AtiyahandN.J.Hitchin, TheGeometryandDynamicsofMagneticMonopoles}; the Killing vector associated to the diagonal ${\rm U}(1)$ is $\xi=\partial/\partial\phi.$ By casting the metric in the form equation~(\ref{eq:self dual Riemman 4- metric}), we establish the identifications
\begin{gather*}
 \frac{1}{V} = \frac{1}{4}\big(\beta^{2}\sin^{2}\theta\cos^{2}\psi+\gamma^{2}\sin^{2}\theta\sin^{2}\psi+\delta^{2}\cos^{2}\theta\big),
\nonumber \\
 \omega = \frac{1}{V}\big(\big(\gamma^{2}-\beta^{2}\big)\sin\psi\cos\psi\sin\theta {\rm d}\theta+\delta^{2}\cos\theta {\rm d}\psi\big),
\end{gather*}
together with
\begin{gather*}
 \gamma_{k^{2}k^{2}} = \frac{\beta^{2}\gamma^{2}\delta^{2}}{V\big(4k^{2}k'^{2}K^{2}\big)^{2}},
\qquad
\gamma_{\theta\theta} = \frac{1}{16}\big[\beta^{2}\gamma^{2}\sin^{2}\theta+\delta^{2}\cos^{2}\theta\big(\beta^{2}\sin^{2}\psi+\gamma^{2}\cos^{2}\psi\big)\big],
\nonumber \\
 \gamma_{\theta\psi} = -\frac{1}{16}\big[\delta^{2}\big(\gamma^{2}-\beta^{2}\big)\cos\theta\sin\theta\cos\psi\sin\psi\big],
\nonumber \\
 \gamma_{\psi\psi} = \frac{1}{16}\delta^{2}\sin^{2}\theta\big(\beta^{2}\cos^{2}\psi+\gamma^{2}\sin^{2}\psi\big).
\end{gather*}
The corresponding K\"ahler potential has a nice simple form as a function of $\theta$, $\psi$, $k^{2}$, namely~\cite{Olivier1991}
\begin{gather*}
\Omega=\frac{\beta\gamma+\gamma\delta+\delta\beta}{4}-J,
\end{gather*}
where
\begin{gather}
J := \frac{1}{8}\big[ (\beta\gamma+\gamma\delta+\delta\beta )-\gamma\delta\sin^{2}\theta\cos^{2}\psi-\delta\beta\sin^{2}\theta\sin^{2}\psi-\beta\gamma\cos^{2}\theta\big].
\label{eq:complex coordinate J}
\end{gather}

On the other hand, Boyer and Finley~\cite{C.P.BoyerandJ.D.FinleyIIIdoi:10.1063/1.525479} studied the Killing vectors in self dual Euclidean Einstein spaces using the formalism of complex $\mathcal{H}$-spaces. They proved that it is always possible to choose complex coordinates such that either
\begin{gather}
\xi=\partial_{p} + \partial_{\bar{p}} \qquad \text{and} \qquad \xi\Omega=0,
\qquad\mbox{or}\qquad
\xi={\rm i}(p\partial_{p} - \bar{p}\partial_{\bar{p}}) \qquad \textrm{and}\qquad \xi\Omega=0,
\label{eq:non antiself dual Killing vector}
\end{gather}
depending on whether the covariant derivative of the Killing vector $\xi$ is purely self-dual or not. In the previous formulae, $\Omega$ is a K\"ahler potential that satisfies the Monge--Amp\`ere equation. Furthermore, they showed how to simplify the equation for $\Omega$ with the help of an appropriate Legendre transformation in both cases. For this purpose, they introduced a pair of complex coordinates $x^{i}\equiv(q,p)$ such that the second alternative in equation~(\ref{eq:non antiself dual Killing vector}) is satisfied, and set
\begin{gather*}
p=\sqrt{r}{\rm e}^{{\rm i}\tilde{\theta}}.
\end{gather*}
In consequence the Killing vector is $\xi=\partial/\partial\tilde{\theta}$ and
\begin{gather*}
\Omega\equiv\Omega(r, q, \bar{q}).
\label{eq:K=0000E4hler potential}
\end{gather*}
By definition of the K\"ahler potential, we have
\begin{gather}
{\rm d}s^{2}=2\Omega_{i\bar{\jmath}}{\rm d}x^{i}{\rm d}x^{\bar{\jmath}},
\label{eq:Definition of the K=0000E4hler potential}
\end{gather}
where $\Omega$ satisfies the Monge--Amp\`ere or first heavenly equation
\begin{gather}
\Omega_{p\bar{p}}\Omega_{q\bar{q}}-\Omega_{p\bar{q}}\Omega_{q\bar{p}}=1.
\label{eq: Deformed Monge-Ampere equation}
\end{gather}
In terms of the variable $r$, the first heavenly equation~(\ref{eq: Deformed Monge-Ampere equation}) becomes
\begin{gather}
\left(r\Omega_{r}\right)_{r}\Omega_{q\bar{q}}-r\Omega_{rq}\Omega_{r\bar{q}}=1.
\label{eq:first Heavenly's equation r=00003Dpp}
\end{gather}
Now, if $J := r\Omega_{r}$ ($J$ is conjugated to $\ln r$ with respect to $\Omega$), and use $(J, q, \bar{q})$ as a new choice of independent variables to rewrite equations~(\ref{eq:Definition of the K=0000E4hler potential}) and~(\ref{eq:first Heavenly's equation r=00003Dpp}), we obtain
\begin{gather*}
\Omega_{q\bar{q}}=r_{J}\left[\frac{r_{q}r_{\bar{q}}}{r (r_{J})^{2}}+1\right],
\end{gather*}
where we used $\Omega_{rq}=-r^{-1} r_{J}^{-1}r_{q}$, $\Omega_{r\bar{q}}=-r^{-1} r_{J}^{-1}r_{\bar{q}}$ and $J_{r}=r_{J}^{-1}$. Equation~(\ref{eq:Definition of the K=0000E4hler potential}) becomes
\begin{gather*}
{\rm d}s^{2}=\frac{1}{2}\left\{ \left(\frac{r}{r_{J}}\right)\left[2{\rm d}\tilde{\theta}+\frac{{\rm i}}{r}\left(r_{q}{\rm d}q-r_{\bar{q}}{\rm d}\bar{q}\right)\right]^{2}+\left(\frac{r_{J}}{r}\right)\big({\rm d}J^{2}+4r{\rm d}q{\rm d}\bar{q}\big)\right\} ,\label{eq:ds in anti-self dual killing vector}
\end{gather*}
so that the line element ${\rm d}s^{2}$ has the same form as equation~(\ref{eq:self dual Riemman 4- metric}); this coordinate frame is referred as the Toda frame~\cite{Bakas:1999wq}. With $J$ given by equation~(\ref{eq:complex coordinate J}), it follows that we have the identifications
\begin{eqnarray}
V = \frac{r_{J}}{2r},
\qquad
{\rm d}\phi+\omega_{i}{\rm d}x^{i} = {\rm d}\tilde{\theta}+\frac{i}{2r}\left(r_{q}{\rm d}q-r_{\bar{q}}{\rm d}\bar{q}\right),
\qquad %\nonumber \\
\gamma_{ij}{\rm d}x^{i}{\rm d}x^{j} = {\rm d}J^{2}+4r{\rm d}q{\rm d}\bar{q}.
\label{changecoords}
\end{eqnarray}

According to Sections~\ref{sec:MOYAL-DEFORMATION-OFTHE FIRST HEAVENLY EQUATION} and~\ref{sec:Integrable-Systems}, a Moyal deformed K\"ahler potential $\widehat{\Omega}$ could be given as a~power series expansion on the noncommutative parameter. If we also impose the condition that the moduli space, which is the AH spacetime, preserves its (anti-)self-dual character under the deformation, we must demand that the rotational Killing symmetry be unchanged; this requirement happens if, and only if, each $\Omega^{(n)}$ in the series expansion of the deformed K\"ahler potential is a function only of $r$, $q$, $\bar{q}$. Therefore, under the assumption $\Omega^{(n)}\equiv\Omega^{(n)}(r, q, \bar{q})$, the original first heavenly equation for the modified K\"ahler potentials $\Omega^{(n)}$ becomes
\begin{gather*}
\sum_{m=0}^s \big\{ \big(r\Omega_{r}^{(m)}\big)\Omega_{q\bar{q}}^{(s-m)}-r\Omega_{rq}^{(m)}\Omega_{r\bar{q}}^{(s-m)} \big\} = 0, \qquad s= 1, 2, \dots.
\end{gather*}
Following the same procedure that for the undeformed case~\cite{Olivier1991}, we start by defining $J^{(n)}:=r\Omega_{r}^{(n)}$ and we use $(J, q, \bar{q})$ as a new choice of independent variables to write $\Omega_{rq}^{(n)}=-r^{-1} J_{r}^{(n)}r_{q}$ and $\Omega_{r\bar{q}}^{(n)}=-r^{-1} J_{r}^{(n)} r_{\bar{q}}$. Using this result and after a lengthy calculation, we obtain the following iterative expression for $\Omega_{q\bar{q}}^{(n)}$
\begin{gather*}
\Omega_{q\bar{q}}^{(n)} = r_{J}J_{r}^{(n)}\big(\Omega_{q\bar{q}}^{(0)}-2r_{J}\big)+\sum_{\substack{s+m=n \\m\neq n, 0 }} r_{J}J_{r}^{(s)}\big(J_{r}^{(m)}r^{-1}r_{q}r_{\bar{q}}-\Omega_{q\bar{q}}^{(m)}\big),
\end{gather*}
where $n\geq1$. The corresponding line element associated to the modified K\"ahler potential $\widehat \Omega$ is then
\begin{gather}
{\rm d}\widehat{S}^{2} = \sum_{n=0}^\infty \theta^{n}J_{r}^{(n)}r_{J}{\rm d}s^{2}+ \bigg[ \sum_{n=1}^\infty -\theta^{n}2r_{J}r_{J}J_{r}^{(n)} + \sum_{n=2}^\infty \theta^{n}\sum_{m=1}^{n-1} r_{J}J_{r}^{(n-m)}
\nonumber\\
\hphantom{{\rm d}\widehat{S}^{2} =}{}
\times \big(J_{r}^{(m)}r^{-1}r_{q}r_{\bar{q}}-\Omega_{q\bar{q}}^{(m)}\big)\bigg] {\rm d}q{\rm d}\bar{q}.\label{eq:dS-deformed}
\end{gather}
It is important to stress that ${\rm d}\widehat{S}^{2}$ possesses the same symmetries as the line element ${\rm d}s^{2}$ of the undeformed K\"ahler potential, namely, they both share the same Killing vector $\xi=\partial/\partial\tilde{\theta}$. In terms of the coordinate pair $(\phi, x^{i})$, we write equation~(\ref{eq:dS-deformed}) as
\begin{gather}
{\rm d}\widehat{S}^{2} = \sum_{n=0}^\infty \theta^{n}J_{r}^{(n)}r_{J}{\rm d}s^{2}+ \sum_{n=0}^\infty \theta^{n}\sum_{m=1}^{n-1} r_{J}J_{r}^{(n-m)}\big(J_{r}^{(m)}r^{-1}r_{q}r_{\bar{q}}+\Omega_{q\bar{q}}^{(m)}\big)
\nonumber \\
\hphantom{{\rm d}\widehat{S}^{2} =}{} \times \frac{1}{4}\big[\gamma_{ij}{\rm d}x^{i}{\rm d}x^{j}-{\rm d}J^{2}\big].\label{eq:dS-deformed2}
\end{gather}
In equation~(\ref{eq:dS-deformed2}), each one of the contributions $J_{r}^{(n)}$ and $\Omega_{q\bar{q}}^{(n)}$ must be expressed in terms of $\phi = \tilde \theta$ and $x^i$. The procedure is straightforward, and we outline it up to first order on $\theta$: first, since both the undeformed AH metric and its K\"ahler potential are known, the metric components $\Omega_{i\bar j}$ are calculated. Then, the first order corrections $\Omega^{(1)}$ and $e^{(1)}$ to the K\"ahler potential and the vierbien are obtained; from them the elements $J^{(1)}$ are also deduced. We obtain the final form by using the change of coordinates in equation~(\ref{changecoords}).

\section{Conclusions}

We analysed the construction of a Moyal deformation of the first heavenly equation that preserves the integrability character of the corresponding K\"ahler potential, as it happens in the standard commutative scenario. For this purpose, we reviewed the Moyal deformation of the first heavenly equation, where the Moyal bracket replaces the standard Poisson bracket; accordingly, a modified potential replaces the commutative K\"ahler potential that satisfies the first heavenly equation. An expression for the modified potential as a series expansion on the noncommutative parameter exists, where each term in this expansion satisfies a partial differential equation~\cite{STRACHAN199263}.

In the standard commutative situation, the K\"ahler potential satisfying the first heavenly equation is integrable. We extended this property to the modified potential by demanding a~set of conditions using the Moyal bracket; these conditions also helped to fix the form of the potential in such a way that it becomes K\"ahler.

We applied these results to the particular case of a K\"ahler potential associated to self-dual vacuum solutions to Einstein's equations, and we analysed the problem of determining each one of the contributions in the series expansion of the modified K\"ahler potential. We obtained then explicit expressions for the deformed vierbein up to first order on the noncommutative parameter. With this information, we obtained two multi-parameter solutions for the K\"ahler potential also to first order.

Finally, we applied this approach to the calculation of the modified K\"ahler potential associated with the AH spacetime, also up to first order on the noncommutative parameter. By extending the procedure of constructing complex coordinates for the AH metric~\cite{Olivier1991}, we obtained thus the modified AH metric in terms of the standard commutative one and the noncommutative contributions to the K\"ahler potential. Taking into account that the AH metric describes the moduli space of interacting magnetic monopoles at low energies, our results aim to incorporate noncommutative effects on these interactions. Furthermore, since the reduction of the AH to the Taub-NUT metric gives the dynamics of two well-separated interacting monopoles at low energies in a classical context, we expect that a deformation induced by noncommutativity would be relevant for this dynamics as well.

It would be interesting also to apply our construction to other spaces, such as the Eguchi--Hanson metric; due to its uncomplicated form, we may find a non-perturbative result for the deformed metric. Considering the unique properties of this metric as a gravitational instanton and its connection with orbifolds and D-branes in asymptotically locally Euclidean spaces, we may find interesting consequences in the context of string theory.

\subsection*{Acknowledgements}
The authors would like to thank the referees for their valuable remarks and suggestions to improve this work. D.~Mart\'inez-Carbajal acknowledges support from Universidad Aut\'onoma Metropolitana (UAM, M\'exico).

\pdfbookmark[1]{References}{ref}
\LastPageEnding

\end{document}